\documentclass[aps,pra,reprint,showpacs,longbibliography]{revtex4-1}

\usepackage{graphicx} 
\usepackage{amsmath}  
\usepackage{comment}  
\usepackage{tabularx}

\begin{document}

\title{Direct magneto-optical compression of an effusive atomic beam for
high-resolution focused ion beam application}

\author{G. ten Haaf}
\thanks{authors contributed equally to this paper}
\author{T.C.H. de Raadt}
\author{G.P. Offermans}
\author{J.F.M. van Rens}
\author{P.H.A. Mutsaers}
\author{E.J.D. Vredenbregt}
\email{e.j.d.vredenbregt@tue.nl}
\author{S.H.W. Wouters}
\thanks{authors contributed equally to this paper}
\affiliation{Department of Applied Physics, Eindhoven University of Technology, P.O. Box 513, 5600 MB Eindhoven, the Netherlands}

\date{\today}

\begin{abstract}
An atomic rubidium beam formed in a 70\,mm long two-dimensional magneto-optical
trap (2D MOT), directly loaded from a collimated Knudsen source, is analyzed
using laser-induced fluorescence. The longitudinal velocity distribution, the
transverse temperature and the flux of the atomic beam are reported. The
equivalent transverse reduced brightness of an ion beam with similar properties
as the atomic beam is calculated because the beam is developed to be
photoionized and applied in a focused ion beam. In a single two-dimensional
magneto-optical trapping step an equivalent transverse reduced brightness of
$(1.0\substack{+0.8\\-0.4})$\,$\times 10^6$\,A/(m$^2$\,sr\,eV) was achieved
with a beam flux equivalent to $(0.6\substack{+0.3\\-0.2})$\,nA. The
temperature of the beam is further reduced with an optical molasses after the
2D MOT. This increased the equivalent brightness to
$(6\substack{+5\\-2})$\,$\times 10^6$\,A/(m$^2$\,sr\,eV). For currents below 10
pA, for which disorder-induced heating can be suppressed, this number is also
a good estimate of the ion beam brightness that can be expected. Such
an ion beam brightness would be a six times improvement over the liquid metal
ion source and could improve the resolution in focused ion beam
nanofabrication.
\end{abstract}

\pacs{37.20.+j,07.77.Ka,89.20.Bb}

\maketitle

\section{Introduction}
Laser cooling and compression is used to intensify atomic beams for use in a
variety of physics experiments such as loading of magneto-optical traps (MOTs),
beam collision studies \cite{Weiner1999} and atom interferometry
\cite{Cronin2009}. A new field of application is the ionization of such cold
atomic beams to create high brightness ion beams which can be applied in
focused ion beams (FIBs) \cite{McClelland2016}. These are table top
instruments in which nanoscale devices can be inspected, by gathering secondary
electrons or ions, and fabricated, by etching and ion beam induced deposition
\cite{Orloff1993,RaffaBook,Gierak2009}. For these applications the most
important figures of merit are the transverse reduced brightness and the energy
spread of the ion beam as these determine the FIBs resolution together with
parameters of the electrostatic lens column. The ion source mostly used in
commercial FIBs for high resolution nanofabrication purposes is currently the
liquid metal ion source (LMIS). This source offers a transverse reduced
brightness of $10^6$ A/(m$^2$\,sr\,eV) \cite{Hagen2008,Gierak2009} and a full
width at half maximum energy spread of 4.5 eV \cite{Bell1988,Gierak2009}. With
this beam quality a LMIS based FIB has a resolution of roughly 5 nm when
operated at 30 kV with a beam current of 1 pA \cite{Orloff1996,Gierak2009}.

Several research groups worldwide have been working on alternative ion sources based on the
field- or photoionization of cold atoms. The first realizations
\cite{Hanssen2008,Reijnders2009} consist of a magneto-optical trap from
which the ions are created and extracted. The ion current, and therefore also
the reduced brightness, from these sources is limited by the slow refilling
rate of the ionization volume inside the MOT. Several proposals
\cite{Freinkman2004,Knuffman2013,Kime2013,Wouters2014} have been made to
overcome this problem by creating a cold atomic beam instead of a MOT, which is
subsequently ionized. 

There are several routes in creating a cold atomic beam. In a so called
2D$^+$ MOT \cite{Dieckmann1998} background vapour atoms are captured in a 2D
MOT in which they are also laser cooled in the third dimension. Through a dark
spot in one of those third dimension laser beams slowly traveling atoms can
escape. This strategy has been proven to produce a flux of $2\times10^{10}$
$^{85}$Rb atoms/s \cite{Chaudhuri2006}. By replacing the third dimension
laser beams by a pair of hollow cooling beams and an additional pushing
beam in the center, the flux can be further optimized. This has recently
been investigated, resulting in a flux of $4\times10^{10}$ Cs atoms/s
\cite{Huang2016}. Without the additional cooling in the third dimension, so with
a pure 2D vapour cell MOT, a flux of $6\times10^{10}$ of faster traveling
$^{87}$Rb atoms/s was produced \cite{Schoser2002}. A so called pyramidal MOT
\cite{Arlt1998} has also been used in the past to create a lower flux of
$4\times10^9$ Cs atoms/s \cite{Camposeo2001}, but with the advantage of only a
single laser cooling beam being used. Instead of capturing the atoms from the
background vapour one can produce similarly valued atomic fluxes by loading a
2D MOT from the transverse direction with an effusive source
\cite{Tiecke2009,Lamporesi2013}. However, in all cold atomic beam experiments
mentioned so far the goal was to produce a large flux of preferably slowly
traveling atoms. In the research presented here, the goal is different.
Here the figure of merit is the brightness of the atomic beam instead of the
flux. Furthermore, for the intended application of transforming the atomic beam
in a high brightness ion beam, the longitudinal velocity of the atoms is less
important which makes longitudinal loading of a 2D MOT with an effusive source
an option. This has already been done in the past, however, in these
experiments a Zeeman slower is usually used in order to slow down the atoms
before entering the 2D MOT \cite{Lison1999,DeGraffenreid2000}, which
drastically increases the size of the apparatus. Furthermore, Tsao et al.
investigated the relative performance of a 2D MOT directly loaded in the
longitudinal direction with a thermal beam of sodium atoms for different
longitunal velocity groups \cite{Tsao1996}.

Here, experimental results are presented of the atomic beam formation in the
atomic beam laser-cooled ion source (ABLIS), in which a 2D MOT is directly
loaded from a collimated Knudsen source \cite{Wouters2016} and used to create a
high brightness $^{85}$Rb atom beam. Extensive simulations of this source, which assume
that all atoms in the beam can be transformed into ions and that
includes the interaction of the ions after ionization, predict that when combined
with a conventional electrostatic focusing column, 1 pA of 30 keV rubidium ions
can be focused to a 1 nm spot \cite{Wouters2014,tenHaaf2014}. This strategy
provides an alternative to the FIB-source developed by Knuffman et al.
\cite{Knuffman2013}, in which a high brightness cesium beam is made by
compressing the beam formed in a 2D$^{+}$-MOT further in a magneto-optical
compressor. Recently, other researchers realized an ion microscope which was
based on the field ionization of a transversely cooled beam of cesium atoms
also originating from a thermal source \cite{Kime2013,Viteau2016}. Here, no
2D trapping or compression was applied and the transverse reduced brightness of
the ion beam was estimated at $2.8\times10^5$ A/(m$^2$\,sr\,eV).

In this paper the quality of the atomic beam  after the 2D
MOT in the ABLIS setup is analyzed by means of laser-induced fluorescence (LIF).
The improvement of the beam quality with an additional optical molasses step is
also explored. The longitudinal velocity distribution, beam flux and transverse
temperature are measured. Also, the equivalent transverse  reduced brightness is
determined, which is defined  as the brightness of an ion beam with similar
temperature and flux density as the atomic beam. Section \ref{sec:methods}
describes the experimental setup in which this is done and the methods used
after which all experimental results are presented in section \ref{sec:results}
and section \ref{sec:Conclusion} presents the conclusions.

\section{Methods\label{sec:methods}}
\begin{figure}[t]
	\includegraphics[width=1.0\linewidth]{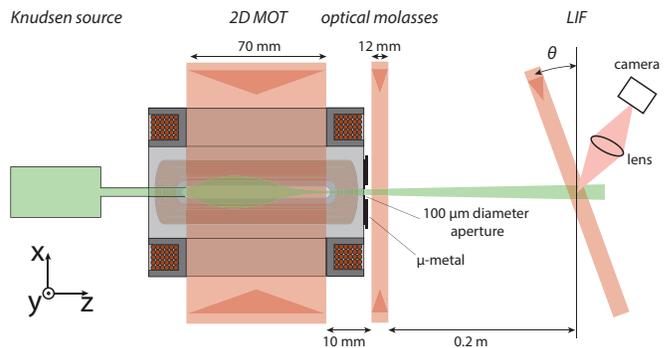}
	\caption{\label{fig:Setup}Overview of the experimental setup (not drawn to
	scale). An atomic rubidium beam effusing from a collimated Knudsen source is
	cooled and compressed by means of two sets of $\sigma^+$/$\sigma^-$ polarized
	laser beams and a compact in-vacuum quadrupole electromagnet with an iron
	core, which form a 2D MOT. A $\mu$-metal plate shields the region after the
	electro-magnet from magnetic fields and an aperture selects the central part
	of the beam. An optical molasses can be created with two sets of laser beams
	with a lin-perp-lin polarization configuration to further reduce the
	transverse temperature of the beam.	The atomic beam is allowed to drift for
	0.2 m where its profile is imaged onto a camera using laser-induced
	fluorescence (LIF). The direction of travel of the probe beam can be altered
	to make an angle $\theta$ with the transverse direction of the atomic beam in
	order to the determine the longitudinal velocity distribution. Note that in
	the actual experiment the atoms travel in the vertical direction.}
\end{figure}

Figure \ref{fig:Setup} shows a schematic picture of the experimental setup. Note
that in the actual experiment the beam travels in the vertical direction since this will also
be the orientation of the source when mounted on a FIB system. As shown, an
atomic rubidium beam from a collimated Knudsen source \cite{Wouters2016} with
temperature $T_s$ effuses into a two-dimensional magneto-optical trap (2D MOT)
\cite{Nellessen1990}. After the 2D MOT the atoms can be cooled to sub-Doppler
temperature with a second set of counter propagating laser beams, forming an
optical molasses. After a 0.2\,m drift, a probe laser beam is used to visualize
the atomic beam by means of laser-induced fluorescence (LIF) which is imaged
onto a camera. Although laser absorption measurements allow very accurate
determination of the beam density \cite{Oxley2016} the choice was made to use
LIF for this is an established technique for beam characterization as well
\cite{Dieckmann1998,Chaudhuri2006,Schoser2002,Tiecke2009,Lison1999,DeGraffenreid2000}.
From the divergence of the atomic beam the temperature is calculated while the
intensity of the LIF signal allows the determination of the flux of the beam.
Both of these calculations require knowledge about the longitudinal velocity
distribution of the atoms in the beam. By placing the probe under an angle with
respect to the atomic beam and scanning its frequency, this distribution is
determined.

The remainder of this methods section is divided into three parts. First the
details of the experimental setup are described. Then the methods to determine the
flux, transverse temperature and equivalent transverse reduced brightness are
introduced and finally the method to measure the longitudinal velocity
distribution will be explained.

\subsection{Experimental setup}
The compact 2D MOT is created with an in-vacuum electromagnet with a pure iron
core, capable of creating a two-dimensional quadrupole field with a magnetic
field gradient $\nabla B$ of 3.8\,T/m. Four identical laser beam expansion
modules (not shown in the figure) generate the required laser fields with a
$\sigma^+$/$\sigma^-$ polarization scheme and a $1/e^2$ diameter of 12\,mm in
the transverse direction (x or y) and 120\,mm in the longitudinal (z)
direction. The peak intensity of each of the four 2D MOT beams is 98\,W/m$^2$
(with a saturation intensity $I_\mathrm{sat}=16.7$ W/m$^2$ \cite{Steck2013}
this gives a saturation parameter of $s=5.9$). The yoke of the magnet has 70\,mm long slots
milled into it to allow the laser beams to reach the center.

Downstream of the 2D MOT laser beams, there is a 10\,mm long drift space in
which the atoms do not see laser light and in which the magnetic field (gradient)
decreases. At the end of the yoke a $\mu$-metal plate is placed to shield the
region behind the yoke from magnetic fields to allow for additional sub-Doppler
cooling. The residual magnetic field gradient after this $\mu$-metal plate was
measured to be 0.05 T/m. Finite element calculations of the magnetic field
in the quadrupole magnet showed that the distortion of the field inside the
quadrupole was less than three percent. An aperture with a 100\,$\mu$m diameter
is also placed after the 2D MOT for beam selection. An imbalance is made in the
currents through the four coils of the quadrupole magnet to steer the atomic
beam through the selection aperture. An optical molasses is created directly
after the selection aperture with two pairs of counter propagating laser beams
with a $1/e^2$ diameter of 12\,mm and a peak intensity of
$\mathrm{1.5\times10^2}$\,W/m$^2$ ($s=4.8$ for $I_\mathrm{sat}=31.8$ W/m$^2$
\cite{Steck2013}). The polarizations of these laser beams are chosen such that
in both directions a lin-perp-lin configuration was achieved.

Laser-induced fluorescence is used to determine the important atomic beam
properties. At a distance of $\Delta z = 0.20$ m from the beam selection
aperture, a probe laser beam propagating over the line $x=y$ is crossed with
the atomic beam. This probe beam has a $1/e^2$ diameter of 11.5\,mm, is linearly
polarized in the z-direction and has a peak intensity of 1.9\,W/m$^2$
($s=0.06$ for $I_\mathrm{sat}=31.8$ W/m$^2$ \cite{Steck2013}) in the
longitudinal velocity distribution measurements and 96\,W/m$^2$ ($s=3.1$ for
$I_\mathrm{sat}=31.8$ W/m$^2$ \cite{Steck2013}) in all other measurements
shown. The fluorescent light emitted by the atoms is imaged onto two cameras;
one looking at the beam in the x-direction and one in the y-direction. This
allows the determination of the temperature of the atomic beam in both
transverse directions. Each measurement series is started with a camera image
with the probe laser far detuned. This image is subtracted from all other
images in the measurement to correct for background scattering.

The laser light for the 2D MOT and optical molasses is generated using a Coherent 899-21
Ti:Saph ring laser. This laser is frequency stabilized at the cross-over
resonance between the $5\,{}^2\text{S}_{1/2}\text{F}=3$ to
$5\,{}^2\text{P}_{3/2}\text{F}'=2,4$ transitions of rubidium-85 in the frequency
modulation spectrum \cite{Bjorklund1983}. An acousto-optical modulator in double
pass configuration is used to shift the laser frequency to the desired
detuning $\delta$ from the $5\,{}^2\text{S}_{1/2}\text{F}=3$ to
$5\,{}^2\text{P}_{3/2}\text{F}'=4$ cooling transition. A resonant
electro-optical modulator is used to generate sidebands at 2918 MHz from the
laser cooling frequency, of which the positive sideband is used for repumping.
Note that the repumping detuning is therefore coupled to the cooling
detuning. Since the frequency difference between the cooling and repumping
transition is 2915 MHz this means that the repumping detuning is approximately
zero when the cooling detuning is set to half a linewidth from the cooling
transition. The cooling and repumping light is coupled into a polarization
maintaining fiber that splits in four and is connected to the optical modules
that shape the laser beams for the 2D MOT and the additional optical molasses.
The light in the probe laser beam is generated using a Toptica DL100-XXL diode
laser. This laser is frequency stabilized by means of a frequency offset lock
that keeps the frequency difference within 100\,Hz of the desired difference
with respect to the Ti:Saph laser frequency. Varying the set point of this
system allows setting the probe laser detuning $\delta_p$ within a range of -90
to 100 MHz.

\subsection{Beam flux, temperature and
brightness\label{sec:beam_flux_temperature_brightness}} 
In this section the equations are presented with which the temperature, flux and
equivalent brightness of the atom beam are extracted from the LIF measurements.
A simple model is set up to describe the transverse density profile of the beam
after the drift from the selection aperture to the probe laser beam. This
profile depends on the velocity distribution of the atoms, so it can be used to
determine the transverse temperature of beam. Also the equations to determine
the atomic flux and density of the beam from the intensity of the LIF signal
are given. Finally the relations between flux, temperature and equivalent
brightness are presented.

After an atom drifts over a length $\Delta z$, its transverse position $x_2$ is
given by $x_2 = x_1 + \Delta z\, v_x/v_z$, in which $x_1$ is the initial
transverse position, $v_x$ is the transverse velocity and $v_z$ is the
longitudinal velocity of the atom. Assuming  $x_1$, $v_x$ and $v_z$ are
uncorrelated and $v_x$ is distributed according a normal distribution with
root-mean-square (rms) width $\sigma_{v_x}$, the mean square size of the
distribution in $x_2$ is given by,
\begin{equation}
	\label{eq:rmsx2}
	\langle x_2^2\rangle =\langle x_1^2\rangle + \left(\Delta z\right)^2
	\sigma_{v_x}^2\langle 1/v_z^2\rangle,
\end{equation}
and the transverse temperature $T_x$ of the beam can be written as
\begin{equation}
	\label{eq:Temp}
	T_x = \frac{m\,\sigma_{v_x}^2}{k_\text{B}} =
	\frac{m}{k_\text{B}}\left(\frac{\widetilde{v_z}}{\Delta z}\right)^2
	\left(\langle x_2^2\rangle - \langle x_1^2\rangle\right),
\end{equation}
in which $m$ is the mass of the atom (taken from ref \cite{Steck2013}), $k_\text{B}$ is Boltzmann's constant and
$\widetilde{v_z} = 1/\sqrt{\langle 1/v_z^2 \rangle}$. Finding the transverse
temperature now relies on measuring $\langle x_2^2\rangle$. Under the
assumption that all the atoms are resonant with the probe laser, the LIF
profile will have the same width as the distribution of $x_2$. This requires
that the Doppler shift of the atoms due to their transverse velocity is smaller
than the linewidth $\Gamma$ \cite{Steck2013} of the transition, i.e. $2 \pi
\sigma_{v_x}/\lambda < \Gamma$ in which $\lambda$ is the wavelength of the
transition. This assumption is valid for transverse beam temperatures lower
than 0.2 K which is easily achieved in the experiment. The initial position
$x_1$ is assumed to be uniformly distributed over the circular selection
aperture. Such a distribution gives $\langle x_1^2\rangle=R/2$, in which $R$ is
the radius of the selection aperture. As will be shown later, this value is
significantly smaller than the spread due to the divergence of the beam due to
which the distribution in $x_2$ resembles a normal distribution very well.
Therefore $\langle x_2^2\rangle$ is found by fitting such a distribution to the
LIF profile.

The scattering rate of LIF photons $R_\text{ph}$ can be found from the
intensity $C$ of the LIF signal. $C$ is acquired by calculating the
area under the normal distribution fitted through the LIF profile and is
expressed in camera counts. Using its value, $R_\text{ph}$ can be calculated
with
\begin{equation}
	R_\text{ph} = \frac{C}{G\, t_\text{c}\, T_\text{geom}\, T_\text{w}\,
	T_\text{f}}\label{eq:scatrate}
\end{equation}
in which $G$ is the number of counts measured by the camera per incident photon,
$t_\text{c}$ is the shutter time of the camera, $T_\text{geom} = \pi
r_\text{l}^2/(4 \pi o^2)$ is the part of the isotropic emission sphere that the
imaging lens with radius $r_\text{l}$ at a distance $o$ from the atomic beam
covers and $T_\text{w}$ and $T_\text{f}$ are the transmission of the vacuum
window and a filter to reduce background light. The beam flux $\Phi$ is
calculated from the scattering rate using
\begin{equation}
	\label{eq:Flux}
	\Phi =  \frac{R_\text{ph}}{\langle t_\text{tr} \rangle
	\rho_{ee}\left(0\right) \, \Gamma},
\end{equation}
in which $\langle t_\text{tr} \rangle$ is the average transfer time through the
imaging volume and $\rho_{ee}\left(\delta_p\right)$ is the excited state
population as a function of the probe laser detuning $\delta_p$, which is given
by
\cite{FootBook}
\begin{equation}
	\label{eq:excited_state_fraction}
	\rho_{ee}\left(\delta_p\right)=\frac{\frac{s_0}{2}}{1+s_0+\left(\frac{2\delta_p}{\Gamma}\right)^2},
\end{equation}
where $s_0=I/I_{sat}$ is the saturation parameter, in which $I$ is the laser
beam intensity and $I_{sat}$ is the saturation intensity \cite{Steck2013}. The
average transfer time through the imaging volume is given by $\langle
t_\text{tr} \rangle = l/\overline{v_z}$, in which $l$ is the longitudinal width
of the imaged volume and $\overline{v_z}=1/\langle 1 / v_z \rangle$. Note that
by writing down equation \ref{eq:Flux} the assumption is made that every photon
that is emitted by an atom in the direction of the imaging lens also reaches the
camera, i.e. reabsorption does not play a role. In section \ref{density}
will be shown that this is a valid approximation.

If all atoms are ionized, the
beam current will be $I=e \Phi$ in which $e$ is the elementary charge. With the
intended application of the beam in a FIB system in mind, the measured flux
will be reported in the unit Ampere as an equivalent ion current throughout
this paper.

Instead of the beam flux, the average beam density $n$ at the selection
aperture, can be calculated using
\begin{equation}
	\label{eq:Density}
	n =  \frac{R_\text{ph}}{\pi R^2\, l\, \rho_{ee}\left(0\right) \, \Gamma}.
\end{equation}
Note that this quantity can be determined without knowing the longitudinal
velocity of the atoms.

Assuming there are no correlations between $x$ and $v_x$ and $y$ and  $v_y$ at
the position of the selection aperture, that the atoms are uniformly
distributed over this circular selection aperture and that $v_x$ and $v_y$ are
distributed according to a normal distribution, the transverse reduced
brightness of an ion beam with equal properties to that of the atomic beam can
be calculated with \cite{Luiten2007}
\begin{equation}
	\label{eq:Brightness}
	B_r = \frac{e\, \Phi}{\pi^2 R^2\, k_\text{B} \sqrt{T_x\,T_y}}.
\end{equation}
Note that if there are correlations between transverse position and velocity at
the selection aperture, the equivalent brightness will be higher than this
calculated value. Furthermore, if the atoms are not uniformly distributed over
the aperture the brightness will be higher as well. This means the calculated
value is a lower limit of the actual peak equivalent brightness. Note that for
presenting the number from equation \ref{eq:Brightness} in the unit of
A/(m$^2$\,sr\,eV) multiplication by a second factor $e$ is required.

Transforming the atomic beam into an ion beam without giving in on the
order of magnitude of transverse reduced brightness is a challenging but not
impossible task. First of all, a high ionization efficiency is desired. One can
estimate \cite{Wouters2014} that in a two-step photoionization process 
in which the excitation light is resonant and the ionization light is above threshold an ionization degree of 80\% can be
achieved within a length of 3 $\mu$m with an ionization laser beam intensity of
$2\times10^{10}$ W/m$^2$. With a laser producing 500 mW of light at the
ionization wavelength of 480 nm this can be realized by focusing the laser beam
to a 1/e$^2$ beam diameter of 8 $\mathrm{\mu}$m. When a build-up cavity is used
a similar intensity can also be achieved over a larger area. Alternatively, one
can use Rydberg excitation and field ionize the atoms
\cite{Kime2013,Viteau2016}. Another point of importance is that after
ionization the ions will strongly interact which can lead to a degradation of
the brightness due to disorder-induced heating. However, in previous work
\cite{tenHaaf2014}, simulations have been presented which show that for
currents below 10 pA these interactions can be suppressed by accelerating the
ions in a large but realistic electric field. Both of these aspects considered,
the equivalent transverse reduced brightness presented in this manuscript also
gives a realistic estimate of the order of magnitude of the transverse reduced
brightness of a 10 pA ion beam that can be made of it. Creation of higher
currents is also possible, but interactions will then limit the transverse
reduced brightness.

An estimate of systematic uncertainties in the experiment was made (see the
Appendix). Due to the many factors involved in the scattering, collection and
conversion of LIF photons, error margins for the temperature
($\substack{+62\\-37}$\,\%), flux ($\substack{+51\\-32}$\,\%), density
($\substack{+48\\-31}$\,\%) and brightness ($\substack{+80\\-38}$\,\%) are
quite substantial.

\subsection{Longitudinal velocity
distribution\label{sec:Longitudinal_velocity_distribution}}

In the calculation of all important beam parameters the value of either
$\overline{v_z}$ or $\widetilde{v_z}^2$ plays an important role. As can be seen
in equations \ref{eq:Temp} and \ref{eq:Flux} the measured temperature scales with
$\widetilde{v_z}^2$ and the measured flux with $\overline{v_z}$. As shown by
equation \ref{eq:Brightness} this means that the measured brightness scales with
with $\overline{v_z}/\widetilde{v_z}^2$. One could assume that the longitudinal
velocity of the atoms in the beam is distributed according to a
Maxwell-Boltzmann distribution with the temperature of the source $T_s$.
However, due to the finite length of the 2D MOT it is to be expected that atoms
with a higher longitudinal velocity are laser-cooled less effectively and thus have a
lower probability of being transmitted through the selection aperture. Therefore
the values of $\overline{v_z}$ and $\widetilde{v_z}^2$ are acquired
experimentally to prevent any errors in the determination of the beam quality
due to a wrongly estimated longitudinal velocity.

The longitudinal velocity distribution $p\left(v_z\right)$ is determined by
looking at the intensity $F\left(\delta_p\right)$ of the fluorescence signal as
a function of the detuning of the probe laser beam that is now oriented so that
it makes an angle $\theta=\left(14.8\pm0.5\right)^{\circ}$ with the atomic beam,
see figure \ref{fig:Setup}. In this way the longitudinal velocity of the atoms will cause a
Doppler shift in the frequency of the laser, so that the effective detuning
becomes $\delta_p-(2 \pi / \lambda) v_z \sin\theta$. Therefore, $F\left(\delta_p\right)$ becomes
dependent on the longitudinal velocity of the atoms. This can be expressed in a
proportionality given by
\begin{equation}
	\label{eq:longitudinal_fluorescence}
	F\left(\delta_p\right)\propto\int_0^\infty\frac{p\left(v_z\right)}{v_z}\rho_{ee}\left(\delta_p-(2 \pi / \lambda) v_z\sin\theta
	\right)\mathrm{d}v_z,
\end{equation}
which explicitly includes all dependence on the longitudinal
velocity of the atoms. It shows that $F\left(\delta_p\right)$ is a convolution
of the line shape of the transition with $p\left(v_z\right)/v_z$, in which the extra
factor of $1/v_z$ compensates for the fact that the transfer time through the
imaged volume is smaller for faster traveling atoms which therefore contribute
less to the fluorescence than slower traveling atoms.

To find $p\left(v_z\right)$, the measured data needs to be deconvoluted from
the linewidth of the transition. Since numerical deconvolution is difficult, the
data was fitted with a test function, that represented
$p\left(v_z\right)/v_z$, which was convoluted with
$\rho_{ee}\left(\delta_p\right)$. For the test function a sixth order polynomial
$P_6\left(v_z\right)$ was chosen, multiplied by a Gaussian distribution
$e^{-\frac{v_z^2}{a^2}}$, in which $a$ is a fitting parameter and
$P_6\left(v_z\right)$ contains the other seven fitting parameters. This means
that after fitting the data, $p\left(v_z\right)$ is calculated with
\begin{equation}
	\label{eq:longitudinal_distribution}
	p\left(v_z\right)=C_n
	v_zP_6\left(v_z\right)e^{-\frac{v_z^2}{a^2}},
\end{equation}
in which $C_n$ is a normalization constant. The resulting distribution is
finally used to calculate the required moments $\overline{v_z}$ and
$\widetilde{v_z}^2$. 

Note that the test function in this fitting routine is not based
on any physical argument. Since the goal was not to find an analytical
expression for the distribution but to deconvolute the measured data from the
linewidth of the transition, a test function with a large number of fitting
parameters was chosen that gives a precise fit to the measured data.
The chosen test function does make sure that the distribution goes to
zero in the limits of $v_z$ to zero and to infinity.

\section{Results\label{sec:results}}

In this section the measurements of the atomic beam parameters are discussed. As
the values of $\overline{v_z}$ and $\widetilde{v_z}^2$ are needed in order to calculate the equivalent
brightness of the beam the longitudinal velocity distribution measurement is
discussed first.

\subsection{Longitudinal velocity distribution}

\begin{figure}
	\centering
	\includegraphics{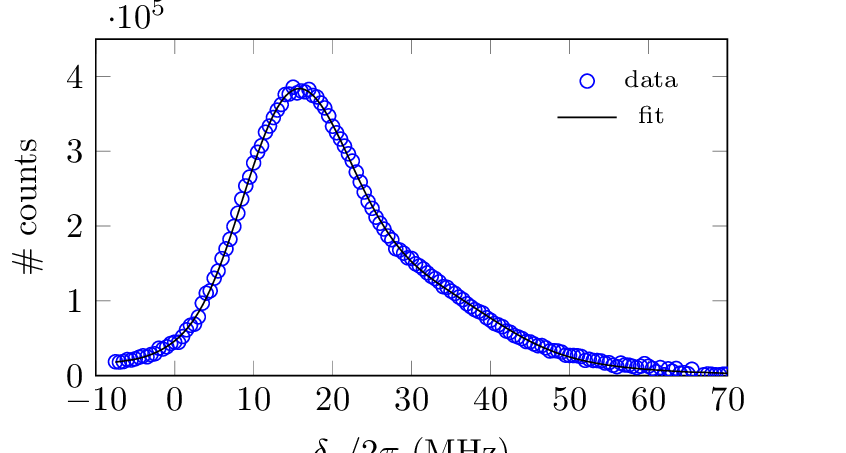}
	\caption{\label{fig:Example_deconvolution} Laser Induced fluorescence signal as
	a function of the laser detuning. The laser made an angle
	$\theta=14.8^{\circ}$ with the transverse direction of the atomic beam. To
	extract the longitudinal velocity distribution the data was fitted with a sixth order polynomial that was multiplied with a
	Gaussian and convoluted with equation \ref{eq:excited_state_fraction}. This
	measurement was performed with $\nabla B=0.94$ T/m,
	$\delta=-1.1\,\Gamma$ and $T_s=433$ K.}
\end{figure}

\begin{figure}
	\centering
	\includegraphics{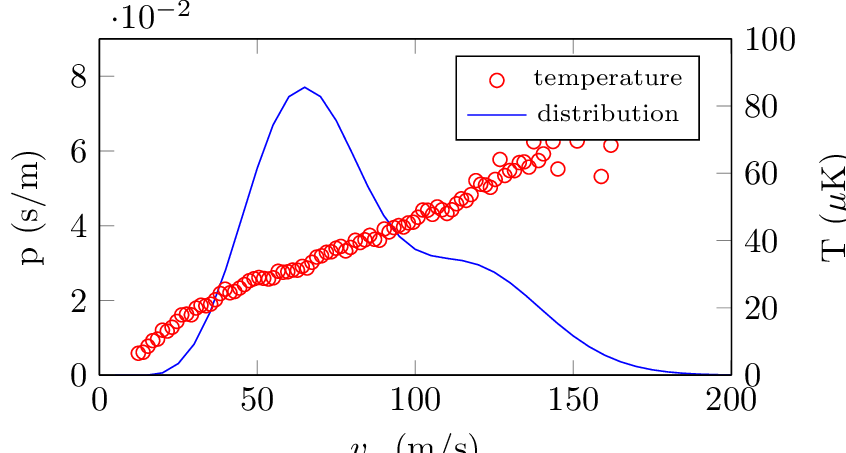}
	\caption{\label{fig:distribution_and_temperature} Longitudinal velocity
	distribution and a plot of the transverse temperature against the longitudinal
	velocity. The longitudinal velocity distribution is found from the data
	plotted in figure \ref{fig:Example_deconvolution}. The
	measurement was performed with $\theta=14.8^{\circ}$, $\nabla B=0.94$ T/m,
	$\delta=-1.1\,\Gamma$ and $T_s=433$ K.}
\end{figure}

Figure \ref{fig:Example_deconvolution} shows an example of a
longitudinal velocity distribution measurement. The measured data was fitted as
described in subsection \ref{sec:Longitudinal_velocity_distribution}. As can be seen the large
number of fitting parameters enables a good fit with the data. The normalized
longitudinal velocity distribution that was obtained from this measurement is
plotted in figure \ref{fig:distribution_and_temperature}. The distribution
yields an average velocity of $\left(83\pm3\right)$ m/s. This is much smaller
than the average velocity of the thermal atoms in the Knudsen source which is
321 m/s. The uncertainty in the average is the uncertainty arising from the
deconvoluting fitting procedure. It is estimated by looking at the spread in the
results when the polynomial in the fitting function contained more or fewer
orders. Apart from this spread there is also a systematic uncertainty on the
value of 2 m/s due to the uncertainties in $\theta_p$ and the absolute value of
$\delta_p$.

As the measured longitudinal velocity distribution in figure
\ref{fig:distribution_and_temperature} results from a deconvolution it is
important to mention that features within a velocity span of approximately
$\Gamma/k\sin\theta=18$ m/s are washed out and not properly represented.
Nevertheless there are some distinct features that are apparent in the
distribution at a larger scale. For example,  after the maximum of the
distribution there appears to be a second bump. This bump is found to be more
pronounced at larger magnetic field gradients. At the highest gradients this
second maximum even became the global maximum in the distribution. The reason
for this bimodal shape of the distribution is not known. An explanation for the
shift to higher average velocities at larger gradients will given below.

An other characteristic of the distribution is that there are no atoms with a
velocity below $\approx20$ m/s. Part of the explanation for this lies in the
fact that the atoms travel through a region of 10 mm with no laser cooling and
compression before being selected by the selection aperture. The slower the
atoms travel longitudinally the larger the divergence of these atoms will be in
this region, which lowers the probability of being transmitted through the
aperture. Furthermore, after the selection aperture, slower traveling atoms
will also have a larger divergence which means that they are more spread out at
the position where they were imaged. Therefore their fluorescence is less
intense and at low enough velocities becomes smaller than the noise level in
the images.

\begin{figure}
	\centering
	\includegraphics{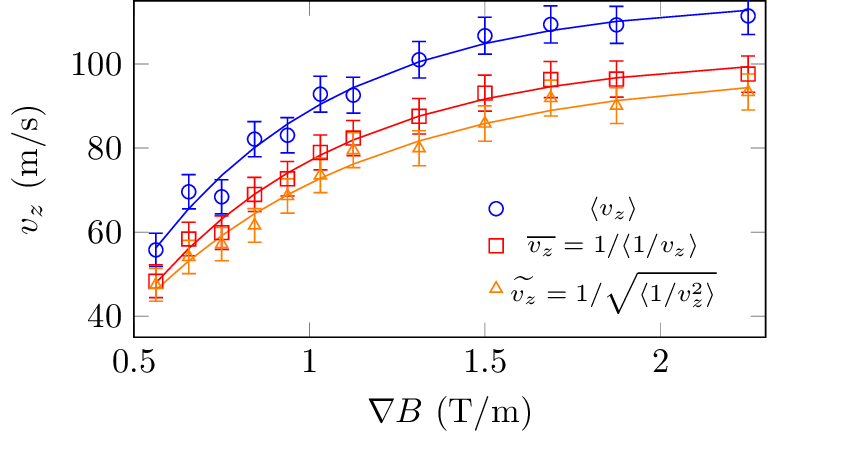}
	\caption{\label{fig:velocity_vs_gradient} Important moments of the longitudinal
	velocity distribution as a function of the magnetic field gradient. The
	error bars show the uncertainty originating from the deconvoluting fitting
	procedure which is estimated at 3 m/s. The fit is performed with an asymptotic
	exponential growth function. This fit function has no physical meaning, but
	serves as a guide to the eye in this figure and is used in further analysis in
	which the moments are used to calculate the flux, temperature and equivalent
	brightness of the beam. The data was measured with $\delta=-1.1\,\Gamma$ and
	$T_s=433$ K.}
\end{figure}

Similar measurements have been performed at different magnetic field gradients. Figure
\ref{fig:velocity_vs_gradient} shows the average velocity and the two
moments required in the calculation of the reduced brightness as a function of
the magnetic field gradient. Below 0.5 T/m the fluorescence signal was not high enough
to perform reliable measurements. Between 0.5 T/m and 1.5 T/m the averages
increase, which can be explained by the fact that by increasing the
magnetic field gradient the spring constant of the 2D MOT becomes larger.
This larger spring constant enables trapping of faster traveling atoms that did
not have enough time to be pushed to the axis at lower gradients. Above a
gradient of 1.5 T/m the averages do not change much. The values of
$\overline{v_z}$ and $\widetilde{v_z}^2$ are used to calculate the flux,
temperature and the equivalent reduced brightness in the next section. Instead
of interpolating between measured values, the data shown in figure
\ref{fig:velocity_vs_gradient} was fitted with an asymptotic exponential growth
function in order to get values for $\overline{v_z}$ and $\widetilde{v_z}^2$ at
magnetic field gradients between 0.5 and 2.5 T/m.

In the derivation of equation \ref{eq:Temp} the assumption was made that no
correlation exists between $v_x$ and $v_z$. In this measurement of the
longitudinal velocity distribution it is possible to check to what extent this
approximation is valid. This is done by looking at the divergence of
different velocity groups in the atomic beam, i.e., by also determining the rms
size of the beam at each detuning and using equation \ref{eq:Temp} to determine
the temperature in which now $\delta_p/k\sin\theta$ was used for the
longitudinal velocity instead of $\widetilde{v_z}$. The results are shown in figure
$\ref{fig:distribution_and_temperature}$. As can be seen, slower traveling
atoms are cooled to a lower temperature than faster traveling atoms, which was
to be expected due to the longer time they spend in the optical molasses. The
approximation made in writing equation \ref{eq:rmsx2} is
$\langle\frac{v_x^2}{v_z^2}\rangle\approx\langle
v_x^2\rangle\langle \frac{1}{v_z^2}\rangle$. However, figure
\ref{fig:velocity_vs_gradient} shows that the average of $v_x^2$ is a linear
function of $v_z$. Substituting this in the approximation above leads to 
$\langle\frac{1}{v_z}\rangle\approx\langle v_z\rangle\langle
\frac{1}{v_z^2}\rangle$. Using the measured velocity distributions, the fraction
of the left- and right-hand side of this approximation can be calculated to
be within 0.74-0.8 over the whole range of measured magnetic field gradients.
This means that the temperature calculated with equation \ref{eq:Temp} yields a
25\%-35\% overestimation of the actual temperature.

\subsection{Beam profiles}
\begin{figure}
	\centering
	\includegraphics{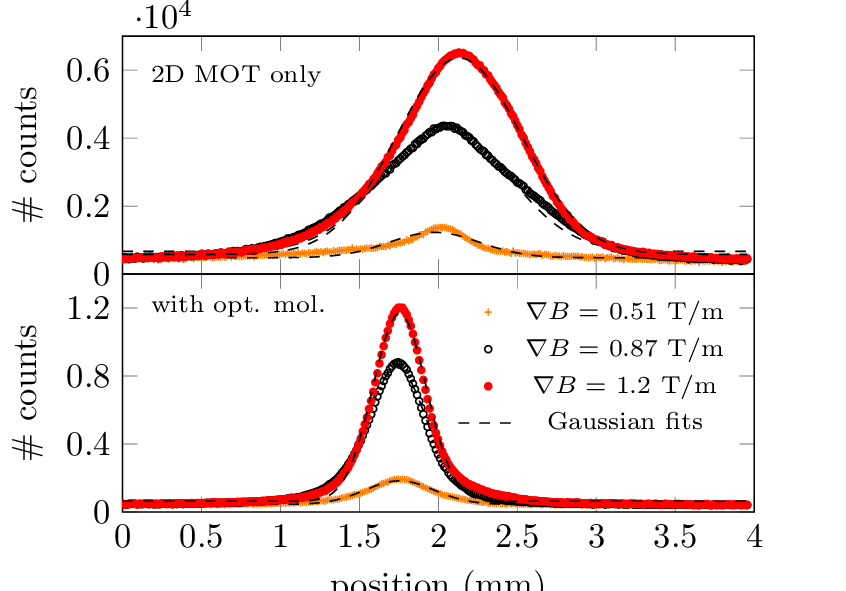}
	\caption{\label{fig:Example} Transverse beam profiles (dots, circles and
	plus signs) after the 2D MOT and a drift of 0.20\,m in the case with (bottom)
	and without (top) optical molasses. The profiles are shown for different magnetic field
	gradients. The figure also shows the Gaussian fits (dashed lines) which were
	used for analysis of the temperature, flux and equivalent brightness of the
	beam. The data was measured with $\delta=-1.1\,\Gamma$ and $T_s=413$ K.}
\end{figure}

Figure \ref{fig:Example} shows several transverse fluorescence profiles of the
atomic beam. In the measurements shown in the top panel, labeled as ``2D MOT
only'', the optical molasses laser beams were turned off while in the
measurements shown in the bottom panel, labeled as ``with opt. mol.'', they were
enabled. As is apparent from the figures, increasing the magnetic field gradient
increases the intensity of the LIF signal. Furthermore, without the optical
molasses, it also leads to a broader profile. With the optical molasses enabled
this does not happen, indicating that the divergence of the beam is indeed
reduced. Careful analysis of the profiles teaches that without optical molasses
the center of the profiles do not overlap for different magnetic field
gradients. The profiles are also slightly asymmetric in this case. Since this
does not happen with the optical molasses enabled, it is attributed to
asymmetries in the 2D MOT, e.g. the imbalance in the currents through the coils
of the magnetic quadrupole that enables the steering of the beam through the
selection aperture. By creating this imbalance, the magnetic axis (where $B=0$)
is overlapped with the selection aperture, but the pointing of the atomic beam
can be altered as well. With the optical molasses enabled, the pointing of the
resulting atomic beam is only determined by the orientation of the optical
molasses laser beams and thus independent of the magnetic field gradient. The
figures also show the Gaussian fits through the profiles, which are used for
further analysis. In the measurements with the optical molasses enabled, the
fits overlap nicely with the data. Due to the asymmetries in the profiles the
overlap is less satisfactory without the optical molasses, but still good
enough for analysis of the beam properties.

\subsection{Effect of cooling laser detuning}
\begin{figure}
	\centering
	\includegraphics{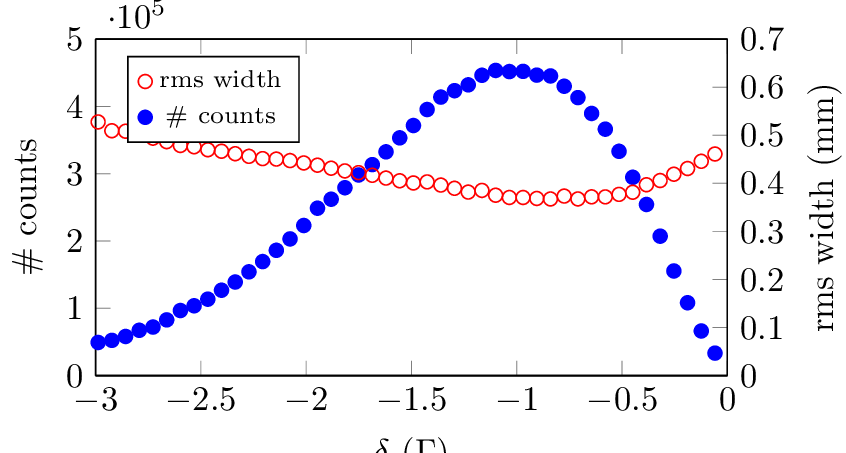}
	\caption{\label{fig:Detuning} Total LIF intensity and rms beam width after
	0.2 m drift as a function of the detuning of the cooling laser.  The data was
	measured with $\nabla B=1.2$ T/m and $T_s=413$ K.}
\end{figure}

An important parameter in the laser cooling and compression process is
the detuning of the cooling laser. A large negative detuning results in a
large capture velocity but a small damping rate of the velocity of the atoms
whereas a small negative detuning results in a small capture velocity but a high
damping rate \cite{FootBook}. To find the optimum in this trade-off, the detuning
was varied and the LIF intensity and profile were monitored. Figure \ref{fig:Detuning} shows
the results of this experiment in which the optical molasses was disabled.
The absolute frequency of the cooling laser was not determined accurately
so the detuning axis was shifted in such a way that no counts where achieved at
$\delta$=0. The results confirm the trade-off between capture velocity and
damping rate. The LIF signal increases from $\delta$=0 to $\delta=$-1.1$\,\Gamma$,
indicating the density at the selection aperture increases. Decreasing the detuning further
from $\delta$=-1.1\,$\Gamma$ reduces the LIF signal again. The rms width also
shows the trade-off: lowering the detuning leads to a decrease, a minimum is
achieved at $\delta$=-0.8\,$\Gamma$ and lowering the detuning even further
increases the rms width again. For a two level atom in an infinitely long laser
cooler (no trapping) with near-zero saturation intensity the lowest temperature is
reached at $\delta$=-$\Gamma$/2 \cite{Chang2014}. Optimum flux and
temperature in the experiment were reached at a different detuning. The
finite length of the 2D MOT, the trapping, the high saturation parameter and
sub-Doppler effects arising due to the multilevel structure of the atom are
possible explanations for this difference. All other measurements shown were
therefore carried out at $\delta$=-1.1\,$\Gamma$.

\subsection{Effect of the magnetic field gradient}
\begin{figure}
	\centering
	\includegraphics{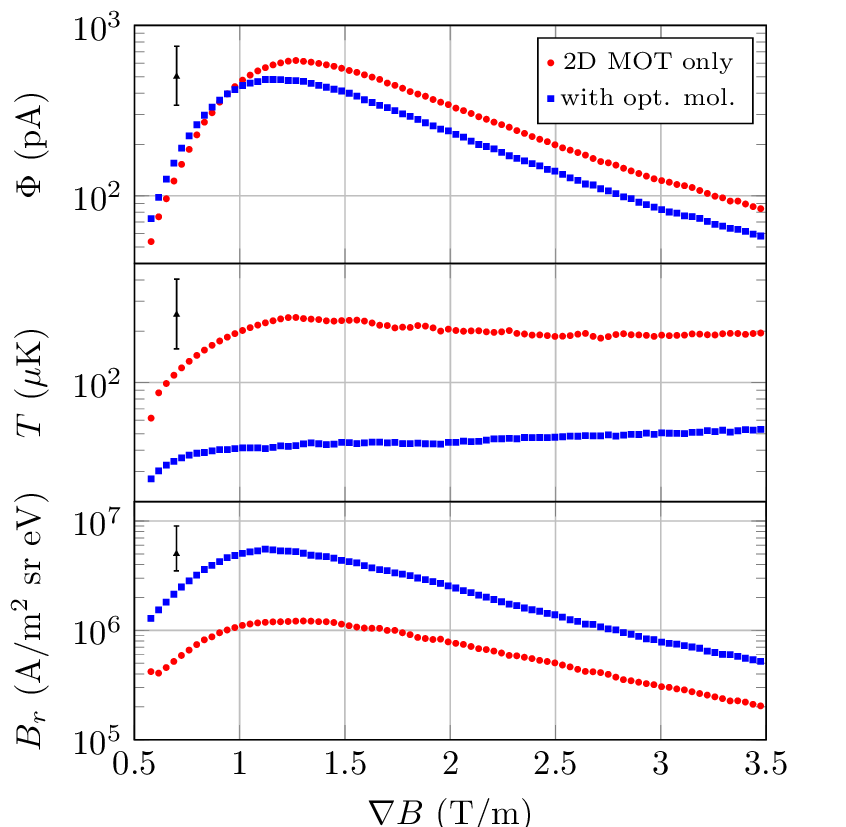}
	\caption{\label{fig:Gradient} Flux, transverse temperature and equivalent
	brightness of the atomic beam as a function of different magnetic field
	gradients. The results with (squares) as well as the results without
	(circles) additional optical molasses are shown. The relative uncertainty
	margin due to systematic errors is indicated by black bars in the upper left
	corner of each graph. In the middle graph the temperatures shown are the
	average of the temperatures in the two transverse directions. The data was
	measured with $T_s$=413\,K and $\delta$=-1.1\,$\Gamma$.}
\end{figure}

Another important parameter determining the properties of the atomic beam is the
magnetic field gradient. Similarly as the detuning, a small magnetic field
gradient leads to a large capture range but a small spring constant
\cite{FootBook}, pulling the atoms towards the magnetic axis. On the other
hand, a high magnetic field gradient leads to a small capture range but a high spring
constant. Furthermore, the presence of a magnetic field can inhibit sub-Doppler
cooling mechanisms that occur due to the $\sigma^+$/$\sigma^-$ laser beam configuration. Figure
\ref{fig:Gradient} shows the flux, transverse temperature and equivalent
brightness as a function of the magnetic field gradient for the experiments with
and without additional optical molasses. The longitudinal velocity distribution was only measured for magnetic field gradients in the range 0.5-2.5 T/m. Since
$\overline{v_z}$ and $\widetilde{v_z}$ are needed in the calculation of all the
parameters shown, they are not determined below 0.5 T/m. Because of the trend
shown in figure \ref{fig:velocity_vs_gradient}, $\overline{v_z}$ and
$\widetilde{v_z}$ are assumed to be constant and equal to their values at 2.5 T/m, for values
above this magnetic gradient.

In the top graph the beam flux through the selection aperture is shown in units
of Ampere, resembling the maximum ion current that can be made from the atomic
beam. As expected, the flux increases when the magnetic field
gradient is raised and decreases again after reaching an optimum. The highest
flux, equivalent to $\left(0.5\substack{+0.3\\-0.2}\right)$nA, was achieved at
a magnetic field gradient of 1.2\,T/m. Since there is no further selection
after the optical molasses section there is no significant difference in flux
in the cases with and without optical molasses.

The middle graph shows the transverse temperature plotted as a function of
the magnetic field gradient. This temperature is the average between the
separately measured temperature in the x- and y-direction. Without optical
molasses the lowest temperature was measured at $\nabla B$=0.6\,T/m and equals
$\left(0.07\substack{+0.04\\-0.03}\right)$\,mK. Furthermore, the trend indicates
even lower temperatures at lower magnetic field gradients. The measured
temperature is below the Doppler temperature of rubidium (0.143\,$\mathrm{mK}$)
suggesting that sub-Doppler cooling effects do play a role, even in the
presence of a magnetic field. Increasing the magnetic field gradient results in
a higher temperature, stabilizing at
$\left(0.2\substack{+0.1\\-0.07}\right)$\,mK which is near the Doppler
temperature. Note that the lowest temperature and highest flux are not achieved
at the same magnetic field gradient. With the optical molasses enabled, the
temperature does not change significantly in the measured range and is equal to
$\left(0.04\substack{+0.02\\-0.01}\right)$\,mK.

The bottom graph shows the equivalent brightness of the atomic beam. Without
the additional optical molasses the highest brightness reads
$(1.0\substack{+0.8\\-0.4})\times10^6$\,A/(m$^2$\,sr\,eV) at a magnetic field
gradient of 1.2\,T/m. Due to the lower temperature the equivalent brightness is
higher with the optical molasses enabled and reads
$(5\substack{+5\\-2})\times10^6$\,A/(m$^2$\,sr\,eV). As was argued at the end of
section \ref{sec:beam_flux_temperature_brightness} the atomic beam can be
ionized and accelerated without major heating below currents of 10 pA.
Therefore, the resulting ion source would be an improvement over the LMIS in
terms of ion beam brightness in this regime.

Table \ref{tab:performance overview} summarizes the performance of the atomic
beam. The flux, temperature, equivalent transverse reduced brightness and
average longitudinal velocity are given at different positions in the setup. As
can be seen only $\approx0.004\%$ of the flux that is leaving the collimated
Knudsen source is transmitted by the selection aperture. This is mostly caused
by the fact that the atoms have a large spread in transverse velocity when
leaving the collimation tube of the Knudsen source, as can be witnessed by the
high transverse temperature at this point stated in table \ref{tab:performance
overview}. Therefore, only a small fraction is being captured by the 2D MOT.
However, by scanning the beam over the selection aperture by displacing the
magnetic axis of the quadrupole it was also found that the FWHM size of the
beam is approximately $3\times$ larger than the diameter of the selection
aperture. This means the total current of the beam is approximately $9\times$
larger before the selection aperture, giving a second explanation for the large
difference in flux before and after the 2D MOT. Note that the flux is of the
same order as most of the atomic beam sources discussed in the introduction
which were aimed at producing a large flux of slow atoms. However, the
difference with those sources is the higher flux density that is achieved here
and therefore the higher brightness. The beam diameter of 100 $\mathrm{\mu}$m
(determined by the selection aperture) is roughly an order of magnitude smaller
as achieved in for example a 2D$^+$-MOT \cite{Dieckmann1998,Chaudhuri2006}. The
two laser cooling sections combined increase the equivalent transverse reduced
brightness with a factor $4\times10^4$ to
$(5\substack{+5\\-2})\times10^6$\,A/(m$^2$\,sr\,eV). This has not been achieved
with a single trapping step when combined with an effusive source. However, a value of
$2\times10^7$\,A/(m$^2$\,sr\,eV) has been achieved by compressing the beam from
a 2D$^{+}$-MOT in a magneto-optical compressor \cite{Knuffman2013} and a value
of $3\times10^7$\,A/(m$^2$\,sr\,eV) has been realized using a setup
incorporating a Zeeman slower \cite{DeGraffenreid2000}.
\begin{table*}
	\centering
	\small
	\caption{Summary of beam parameters at several points in the setup. All values
	are measured/calculated at a Knudsen source temperature of 413 K. The values
	measured after the 2D MOT and optical molasses are stated for the magnetic
	field gradient at which the equivalent reduced
	brightness was maximized.\label{tab:performance overview}}
	\begin{tabularx}{\linewidth}{X|c|c|c|c}
		\hline
		\hline
		Position&$\Phi$ (1/s)&$T$ (mK)&$B_r$ (A/m$^2$/sr/eV)&$\langle v_z\rangle$ (m/s)\\ \hline
		After collimated Knudsen
		source&$\left(9\pm4\right)\times10^{13}$\textsuperscript{\textdagger}&$\left(94\pm10\right)\times10^3$\textsuperscript{\textdagger}&$\left(140\pm40\right)$\textsuperscript{\textdagger}&321\textsuperscript{\textdagger\textdagger}\\
		After 2D
		MOT&$\left(4\substack{+2\\-1}\right)\times10^9$&$\left(0.2\substack{+0.1\\-0.07}\right)$&$(1.0\substack{+0.8\\-0.4})\times10^6$&$\left(93\pm2\right)$\\
		After optical
		molasses&$\left(3\substack{+2\\-1}\right)\times10^9$&$\left(0.04\substack{+0.02\\-0.01}\right)$&$(5\substack{+5\\-2})\times10^6$&$\left(93\pm2\right)$\\
		\hline
		\hline
		\multicolumn{5}{l}{\textsuperscript{\textdagger}\footnotesize{Value was adapted from
		previous measurements discussed in Ref. \cite{Wouters2016}}}\\
		\multicolumn{5}{l}{\textsuperscript{\textdagger\textdagger}\footnotesize{Value was not measured
		but calculated from Knudsen source temperature}}
	\end{tabularx}
\end{table*}

\subsection{Beam density vs. source temperature \label{density}}
The last parameter that was varied is the temperature of the Knudsen source.
Under the assumption of no collisions or other density limiting effects,
increasing the temperature of the source would lead to an increase in flux
according to equations 1 and 12 from \cite{Wouters2014}. However it is known
from experiments on 3D MOTs that at high densities and intense
resonant illumination, inelastic collisions between ground and excited state
atoms \cite{Wallace1992} and attenuation and radiation trapping effects
\cite{Walker1990} will limit the achievable density. Therefore an experiment
was performed in which the temperature of the Knudsen source was varied.

\begin{figure}
	\centering
	\includegraphics{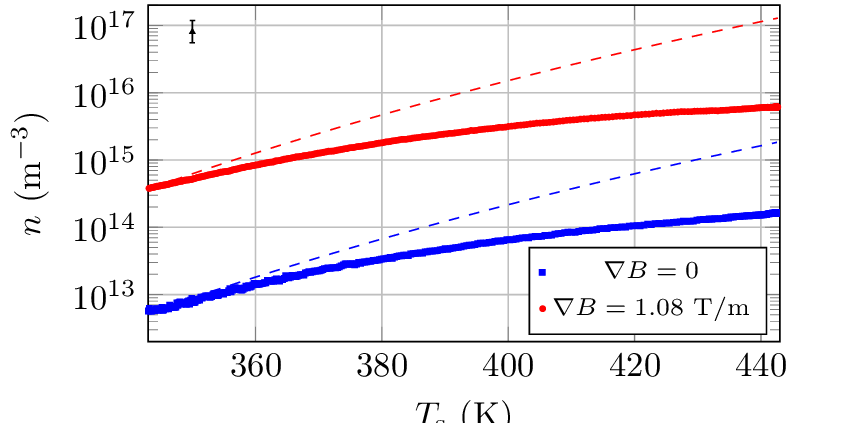}
	\caption{\label{fig:Source} Atomic beam density after the 2D MOT as a function
	of the source temperature. The relative uncertainty margin due to systematic
	errors is indicated with a black bar in the upper left corner.
	Two measurements are shown: one in which the beam is
	only cooled ($\nabla$B=0\,T/m, the blue open circles) and one in
	which it is also compressed ($\nabla$B=1.1\,T/m, red dots). In all measurements
	the detuning of the cooling laser was set to -1.1\,$\Gamma$ and the detuning of the
	probe laser was set to 0\,MHz. As a reference, the dashed lines show a
	scaling of the first data point with the theoretical flux comming from the Knudsen
	source under the assumption of no collisions inside the collimation tube
	\cite{Wouters2016}.}
\end{figure}

Figure \ref{fig:Source} shows the beam density as a function of source
temperature in two cases: with the magnetic field gradient set to the optimal
value of 1.1\,T/m and without any magnetic field. The figure also shows a
scaling law that scales the first data point of both measurements with the flux
coming from the Knudsen source \cite{Wouters2016}. Although the beam density
does increase with increasing source temperature, figure \ref{fig:Source} shows
that the scaling law only holds for the lowest temperatures and in the case of
no magnetic field gradient. At the highest temperature the measurement and the
scaling are off by a factor 11 in the case without a magnetic gradient and a
factor 21 in the case with a magnetic gradient. This deviation from the scaling
is attributed to three effects. First of all, in the determination of total
flux of atoms effusing from the collimated Knudsen source, see ref
\cite{Wouters2014}, it was observed that at high temperatures the flux was
lower than expected from the model. At 433\,K the difference is a factor 2,
partly explaining the difference in this experiment. In the measurement of the
flux from the collimated Knudsen source it was also observed that the
transverse velocity distribution of the atoms was broadened due to collisions
in the collimation tube. This broadening reduces the centerline intensity and
thus reduces the fraction of atoms that can be captured by the 2D MOT. Between
343 and 433\,K the width of the velocity distribution increases by a factor 2
suggesting that the capturable fraction decreases by a factor 4. The effects of
a lower flux and a broader transverse velocity distribution from the collimated
Knudsen source can explain most of the difference between the results from the
measurement without a magnetic gradient and the scaling law. The additional
limiting of density in the experiment in which the magnetic field is enabled
can not be explained by effects caused by the collimated Knudsen source.
However as can be seen from the results, at high temperatures the beam density
approaches 10$^{16}$\,m$^{-3}$ which is the density in which collisions
between excited and ground state atoms and radiation trapping effects start to
play a role in MOTs\cite{Wallace1992,Walker1990,Ketterle1993}. More advanced laser
cooling schemes using a repumper laser beam that has a dark spot in the very
center of the atomic beam may allow overcoming this limitation
\cite{Ketterle1993}. Because of the added complexity this was not pursued in
this experiment.

Ultimately, a beam density of
$\left(6\substack{+3\\-2}\right)$$\times$10$^{15}$\,m$^{-3}$ was reached. This
translates in a beam flux equivalent of $\left(0.6\substack{+0.3\\-0.2}\right)$
nA through the selection aperture and an equivalent brightness of
$(6\substack{+5\\-2})\times10^6$\,A/(m$^2$\,sr\,eV). The current Knudsen source
does not allow for higher temperatures than $\approx$443\,K. However, linear
extrapolation of the measurement teaches that only a 20\,\% increase in
density, and thus brightness, can be achieved when the source temperature is
raised by 20\,K. Since part of the deviation from the scaling law is caused by
collisions inside the collimation tube of the Knudsen source a slightly higher
flux can be expected when using a Knudsen source collimated by an aperture
instead of a tube or the more complex candlestick oven design
\cite{Walkiewicz2000}. However, ultimately the density will be limited by the
earlier mentioned effects.

With the maximum beam density measured and the cross section for absorption on
resonance \cite{Steck2013} the optical density of the beam can be calculated to
be $9\times10^{-2}$ at the position of the selection aperture. This means
that if the LIF experiment was performed at the aperture 91\% of the emitted
photons would be transmitted through the atomic beam. The resulting correction for (re-)absorption is smaller than the error margins.


\section{Conclusion\label{sec:Conclusion}}

The properties of an atomic rubidium beam resulting from 2D magneto-optical
trapping of effusive atoms from a collimated Knudsen source are
evaluated. Laser-induced fluorescence is used to image the beam and determine
its flux and transverse temperature. As for both of these parameters knowledge
of the longitudinal velocity of the atoms was required, the longitudinal
velocity distribution was measured as well. The average longitudinal velocity
was found to be dependent on the magnetic field gradient in the 2D MOT and had a
value of 50-100 m/s. As the beam is intended to be photoionized and applied as
a source for a focused ion beam, its flux is expressed in units of current and
also the equivalent reduced brightness is calculated. In a single
2D magneto-optical trapping step the maximum equivalent beam current found was
$\left(0.6\substack{+0.3\\-0.2}\right)$ nA and the transverse temperature was
$\left(0.2\substack{+0.1\\-0.07}\right)$\, mK. Together these values combine
into an equivalent transverse reduced brightness of
$\left(1\substack{+0.8\\-0.4}\right)\times$10$^6$\,A/(m$^2$\,sr\,eV). With an
additional optical molasses step this value was increased to
$\left(6\substack{+5\\-2}\right)\times10^6$\,A/(m$^2$\,sr\,eV) by lowering the
transverse temperature of the beam. When ionized, this would be a $6\times$
improvement over the brightness of the liquid metal ion source, $300\times$
larger than any MOT based ion source \cite{Hanssen2008,Debernardi2012} and
similar to the estimated brightness from the 2D$^+$-MOT based source by
Knuffman et. al \cite{Knuffman2013}.

In future research photoionization of the atomic beam will be investigated.
To reach an ionization degree of 80\%, an ionization laser intensity of of
$2\times10^{10}$ $\mathrm{W/m^2}$ is needed. In order to reach this intensity
over a large cross sectional area of the beam a build-up cavity will be used to
enhance the power of a commercially available laser system. Furthermore,
previous work \cite{tenHaaf2014} predicted that by immediately accelerating the
ions in a sufficiently large but realistic electric field the transverse
reduced brightness of the beam can be conserved at currents below 10 pA.
Therefore, nanometer sized waists are expected when focusing a 30 keV beam
containing 1 pA.

\begin{acknowledgments}
This research is supported by the Dutch Technology Foundation STW, which is part
of the Netherlands Organisation for Scientific Research (NWO), and which is
partly funded by the Ministry of Economic Affairs. The research was also
supported by FEI Company, Pulsar Physics and Coherent Inc. We would like to
thank Bas van der Geer and Eddy Rietman for their design work on the magnetic
quadrupole, and Eddy Rietman, Harry van Doorn and Iman Koole for their
technical support.
\end{acknowledgments}

\appendix*
\section{Uncertainty analysis}
In linear uncertainty analysis the dependence of the end results on its
parameters is linearized. This gives wrong results if the relative errors,
$\Delta i/i$ are large. Therefore the upper and lower (systematic) value for
the flux, temperature and brightness were calculated differently by filling in
the parameters plus or minus their uncertainty margin in such a way that the
maximal or minimal value for the flux, temperature or brightness was found. The
complete equation used for finding the transverse temperature from the
experiment is given by
\begin{equation}
	T_x = \frac{m}{k_\text{B}} \cdot \frac{\widetilde{v_z}^2}{(\Delta z)^2} \left( \frac{o^2\, l_\text{px}^2}{\, b^2} \cdot \sigma_\text{px}^2 - (R/2)^2 \right),
	\label{eq:temp2}
\end{equation}
in which $\sigma_\text{px}$ is the root-mean-square width of the fitted
normal distribution in units of camera pixels and $l_\text{px}$ is the
width of a single pixel. The complete equation for the flux was find by
combining equations \ref{eq:scatrate}-\ref{eq:excited_state_fraction}, resulting
in
\begin{equation}
	\Phi = \frac{4}{\Gamma}\cdot \frac{b\, \overline{v_z}\, o\,C}{l_\text{img}\,
	\rho_{ee}\left(\delta_p\right)\, r_l^2\, t_\text{c}\, T_w\, T_f\,
	G},\label{eq:flux2}
\end{equation}
in which $l$ has been substituted with $l=l_\text{img}\frac{o}{b}$, in which
$l_\text{img}$ is the longitudinal size of the image. Using the same
substitution and equation \ref{eq:Density} for the density results in,
\begin{equation}
	n = \frac{4}{\pi\Gamma}\cdot \frac{b\,o\,C}{R^2l_\text{img}\,
	\rho_{ee}\left(\delta_p\right)\, r_l^2\, t_\text{c}\, T_w\, T_f\,
	G}.
\end{equation}
The complete equation for the equivalent brightness can be found by combining equation \ref{eq:Brightness}, \ref{eq:temp2} and \ref{eq:flux2} resulting in
\begin{equation}
	\begin{split}
	B_r = \frac{4\, e}{m\, \pi^2\, \Gamma} \cdot \frac{b^3\, o\, (\Delta z)^2}{R^2\, l_\text{img}\, \rho_{ee}\left(\delta_p\right)\, r_l^2\, t_\text{c}\, T_w\, T_f\, G}\\
	\frac{C\, \overline{v_z}}{\widetilde{v_z}^2\, \left( o^2\,l_\text{px}^2\,\sigma_\text{px}^2 - b^2\,(R/2)^2 \right)}.
	\end{split}
\end{equation}
Typical values of all parameters used, are given in table \ref{tab:Errors}
together with the uncertainties in them. With these values the relative uncertainties are +62\,\% and -37\,\% for the
temperature, +51\,\% and -32\,\% for the flux, +48\,\% and -31\,\% for the
density and +80\,\% and -30\,\% for the equivalent brightness.

\begin{table*}
	\centering
	\small
	\caption{\label{tab:Errors}Typical experimental parameters and their systematic
	uncertainties}
	\begin{tabular}{l|r|r}
		\hline
		\hline
		Parameters (unit) & Symbol & Value $\pm$ error \\
		\hline 
		Probe laser detuning (MHz) & $\delta_p$ & $0 \pm 1$ \\
		Probe laser beam intensity (W/m$^2$) & $I$ & $96 \pm 10$ \\
		Probe laser beam angle for $v_z$ measurement ($^\text{o}$) & $\theta$ & $14.8 \pm 0.5$ \\
		Saturation parameter for linear polarized light (W/m$^2$) & $I_\text{sat}$ & 31.8 + 7.2 - 0.0 \\
		Drift distance (m) & $\Delta z$ & $0.20 \pm 0.02$ \\
		Camera shutter time (s) & $t_\text{c}$ & $1.000 \pm 0.001$ \\
		Camera gain (counts/photon) & $G$ & $2.08 \pm 0.08$ \\
		Camera pixel size ($\mu$m) & $l_\text{px}$ & $5.86$ \\
		Longitudinal image size (mm) & $l_\text{img}$ & $1.172$ \\
		Probe window transmission (\%) & $T_\text{w}$ & $90 \pm 2$ \\
		Filter transmission (\%) & $T_f$ & $89 \pm 1$ \\
		Selection aperture radius ($\mu$m) & $R$ & $50 \pm 1$ \\
		Lens object distance (m) & $o$ & $0.237 \pm 0.005$ \\
		Lens image distance (m) & $b$ & $0.105 \pm 0.005$ \\
		Lens aperture radius (mm) & $r_l$ & $6 \pm 0.5$ \\
		Resulting inverse averaged longitudinal velocity (m/s) & $\overline{v_z}$ & $82 \pm 3$\\
		Resulting inverse squared averaged longitudinal velocity (m/s) & $\widetilde{v_z}$ &$79\pm 3$ \\
		\hline
		\hline
	\end{tabular}
\end{table*}

\bibliography{MOC_Results}

\end{document}